# Synthesis of MAX Phases Nb$_2$CuC and Ti$_2$(Al$_{0.1}$Cu$_{0.9}$)N by A-site Replacement Reaction in Molten Salts

*Haoming Ding$^{a,b\dagger}$, Youbing Li $^{b,c\dagger}$, Jun Lu $^d$, Kan Luo$^b$, Ke Chen $^b$, Mian Li $^b$, Per O. Å. Persson $^d$, Lars Hultman $^d$, Per Eklund $^d$ , Shiyu Du $^b$, Zhengren Huang $^b$, Zhifang Chai $^b$, Hongjie Wang $^a$, Ping Huang $^a$, Qing Huang $^{b*}$*

*$^a$State Key Laboratory for Mechanical Behavior of Materials, Xi'an Jiaotong University, Xi'an, Shaanxi, 710049, China.*
*$^b$Engineering Laboratory of Advanced Energy Materials, Ningbo Institute of Materials Technology and Engineering, Chinese Academy of Sciences, Ningbo, Zhejiang, 315201, China.*
*$^c$University of Chinese Academy of Sciences, 19 A Yuquan Rd, Shijingshan District, Beijing 100049, China.*
*$^d$Department of Physics, Chemistry, and Biology (IFM), Linköping University, S-58183 Linköping, Sweden.*

**Abstract**

New MAX phases Ti$_2$(Al$_x$Cu$_{1-x}$)N and Nb$_2$CuC were synthesized by A-site replacement by reacting Ti$_2$AlN and Nb$_2$AlC, respectively, with CuCl$_2$ or CuI molten salt. X-ray diffraction, scanning electron microscopy, and atomically-resolved scanning transmission electron microscopy showed complete A-site replacement in Nb$_2$AlC, which lead to formation of Nb$_2$CuC. However, the replacement of Al in Ti$_2$AlN phase was only close to complete at Ti$_2$(Al$_{0.1}$Cu$_{0.9}$)N. Density-functional theory calculations corroborated the structural stability of Nb$_2$CuC and Ti$_2$CuN phases. Moreover, the calculated cleavage energy in these Cu-containing MAX phases are weaker than in their Al-containing counterparts, indicating that they are precursor candidates for MXene derivation.

**KEYWORDS:** MAX phase; replacement reaction; copper; density functional theory

# Introduction

The MAX phases constitute a family of ternary compounds with a hexagonal structure (space group $P_{63}/mmc$, *194*) and a molecular formula of $M_{n+1}AX_n$, where M is an early transition metal, A mainly comes from groups 13-16, X is carbon and/or nitrogen, and n=1-3 [1,2]. The MAX phases have potential applications in high-temperature electrodes, components with resistance to friction and wear, structural material in nuclear fuel cladding, and as precursor material for two-dimensional MXene [2-4]. By now, more than 80 members of ternary MAX compositions have been discovered [5]. Recent studies have also demonstrated that the A-site element in MAX phases can be a late transition metal (e.g., Au, Ir, Zn, Fe and Cu) [6-13]. Transition metals have distinct properties different from other A-group elements due to their large *d* electron orbits. If late-transition metal elements can be introduced into the A layer of the MAX phase through a replacement reaction, there would be further prospects for tailoring the functionality of MAX phases.

In 2017, $Ti_3AuC_2$ and $Ti_3Au_2C_2$ were synthesized by replacing Si with Au in $Ti_3SiC_2$, and $Ti_3IrC_2$ was identified by replacing Au with Ir in obtained $Ti_3Au_2C_2$ [7]. Recently, our group reported a series of new Zn-containing MAX phases ($Ti_3ZnC_2$, $Ti_2ZnC$, $Ti_2ZnN$, and $V_2ZnC$) obtained by a replacement reaction between MAX phase precursors and $ZnCl_2$ molten salt [10]. In these phases, Zn atoms occupy the original Al position at the A site in the MAX phase structure. The key merit of this A-site replacement strategy is the prevention of competitive phases (such as M-Zn alloys) that can have lower Gibbs free energies than these new MAX phases and would thus be thermodynamically favored. In this synthesis methodology, the redox reaction between Al and $Zn^{2+}$ and simultaneous evaporation of $AlCl_3$ accounts for the main driving force. Since $Cu^{2+}$ cations have higher oxidation potential than $Zn^{2+}$ cations, Cu atoms have also been incorporated into Ti$_3$AlC$_2$, partially occupying the Al in resultant Ti$_3$(Al$_{1/3}$Cu$_{2/3}$)C$_2$ MAX phase through a similar replacement approach [14]. The partial substitution behavior of Cu in Ti$_3$(Al$_{1/3}$Cu$_{2/3}$)C$_2$ was explained according to the Cu-Al binary phase diagram, in which intermediate Cu-Al alloys are in equilibrium with Al metal below the solidus line. When part of the Al is consumed in a redox reaction and driven out in the form of AlCl$_3$, Cu and residual Al atoms occupy the A layer of as-formed Ti$_3$(Al$_{1/3}$Cu$_{2/3}$)C$_2$. Although binary phase diagrams (Au-Si, Al-Zn, Al-Cu) have been used to describe the substitution/replacement behavior in these new MAX phases, the atom stacking or mutual atomic interaction in two-dimensional single-atomic A layer should be different from that in three-dimensional materials.

In order to expand this replacement strategy to other novel MAX phases, studies are needed on the incorporation of Cu into a range of MAX phases. Here, Nb$_2$CuC and Ti$_2$(Al$_{0.1}$Cu$_{0.9}$)N were synthesized by Cu-substitution for Al in Nb$_2$AlC and Ti$_2$AlN phases by reaction with CuI and CuCl$_2$ molten salts.

## Experimental details

### Preparation of Ti$_2$AlN and Nb$_2$AlC

As in previous work, TiN/Ti/Al/NaCl/KCl powder mixture with a mole ratio of 1: 1: 1: 4: 4, and NbC/Nb/Al powder mixture with mole ratio of 1:1:1 was sintered in order to synthesize Ti$_2$AlN and Nb$_2$AlC MAX phase powders. For more details, refer to the Supplementary Information.

### Preparation of Ti$_2$(Al$_x$Cu$_{1-x}$)N and Nb$_2$CuC

The Ti$_2$AlN powders were mixed with CuCl$_2$ in stoichiometric molar ratios of 2:3 for Ti$_2$(Al$_x$Cu$_{1-x}$)N. The Nb$_2$AlC and CuI (molar ratio=1:3) were used as the starting material to synthesize Nb$_2$CuC. The material mixtures was heated in tube furnace to 600°C at a rate of 2°C/min for 7 h under the protection of argon, then cooled down to room temperature at a rate of 5°C/min. Ammonium persulfate solution was used to remove the residual Cu in the reaction process. Finally, the product was filtered, washed, and dried at 50°C.

**Characterization and density functional theory calculations**

The phase composition of the samples was analyzed by X-ray diffraction (XRD) with Cu Kα radiation. The microstructure and chemical composition were obtained in scanning electron microscopy with an energy-dispersive spectrometer (EDS). Atomically-resolved structural analysis was also carried out by high-resolution scanning transmission electron microscopy (HRSTEM) capable of high angle annular dark field (HAADF) imaging and EDS

Density functional theory (DFT) calculations were in the in the CASTEP code [15,16], using the generalized gradient approximation (GGA) as implemented in the Perdew-Breke-Ernzerhof (PBE) functional[17,18]. Phonon calculations were carried out to evaluate the dynamical stability using the finite displacement approach, as implemented in CASTEP[19,20]. The equation $E=(E_{broken}-E_{bulk})/S$ [10] was adopted to calculate the cleavage energy E. In this equation, $E_{bulk}$ and $E_{broken}$ represent the total energies of bulk MAX and the cleaving structures with a 10 Å vacuum separation in the corresponding M and A atomic layers, while $S$ is the cross-sectional surface area of the MAX phase materials.

More details are provided in the Supplementary Information.

**Results and discussion**

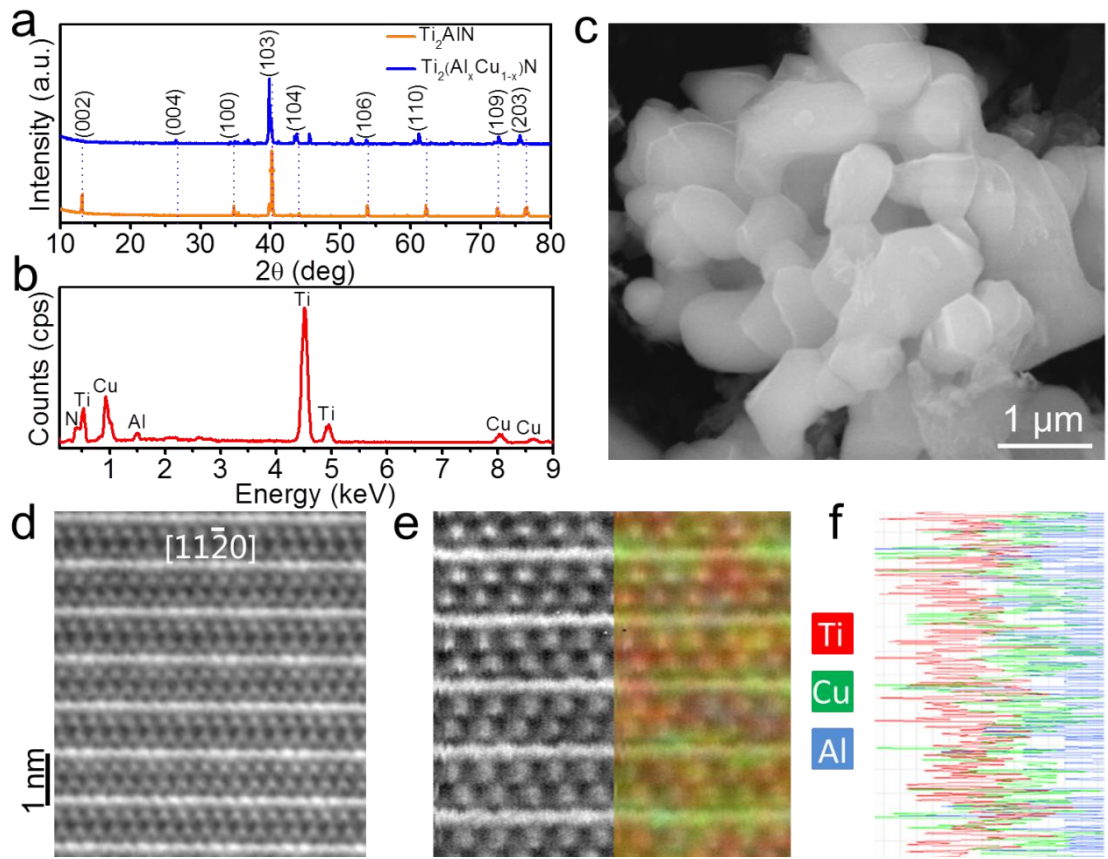

**Figure 1.** (a) XRD patterns of the $Ti_2AlN$ and the $Ti_2(Al_{0.1}Cu_{0.9})N$ obtained from the reaction of $Ti_2AlN$ and $CuCl_2$. (b) SEM image of the $Ti_2(Al_{0.1}Cu_{0.9})N$ powder and (c) corresponding EDS spectrum. (d) High-resolution (HR)-STEM images of $Ti_2(Al_{0.1}Cu_{0.9})N$ showing atomic positions along $[11\bar{2}0]$ direction. (e) Element mapping in STEM-EDS mode and (f) corresponding line–scan of Ti-Kα (red), Cu-Kα (green) and Al-Kα (blue) signals, respectively.

The Cu-substituted MAX phase was synthesized by replacement reaction between $Ti_2AlN$ and $CuCl_2$ molten salt at 600°C. Figure 1(a) shows XRD pattern of the resulting material. Compared to $Ti_2AlN$, the (103), (104), and (106) peaks are shifted towards lower angles. The (002) peak is almost vanished and the (004) peak becomes strong, indicating a change in periodic symmetry along the *c* axis and corresponding change in structure factor [10]. The EDS measurement (Figure 1(b)) on the resultant particles provide a semi-quantitative estimation of the atomic ratios, with of Ti:(Cu+Al) and Cu:Al being 2:1 and 9:1, respectively. Moreover, elemental mapping indicate that Ti, Cu, Al, and N are uniformly distributed in their respective atomic layers of the MAX structure (See Figure S(1)). Since the content of Al in the final product was low, we calculated the lattice parameters to study the effect of Cu incorporation assuming full replacement. The result shows that $a$=3.026 Å and $c$=13.402 Å in the Cu-incorporated MAX phase as compared with $a$=2.999 Å and $c$=13.650 Å of $Ti_2AlN$ [21], indicating in-plane expansion with corresponding reduction in the $c$ axis after replacement. Figure 1(c) also shows that the morphology of the resulting particles is similar to that of the parent $Ti_2AlN$ particles in size and shape (shown in Figure S(2)).

Figure 1(d-f) show STEM images of the resulting phase. The atomic positions perpendicular to the $[11\bar{2}0]$ zone axis is shown in Figure 1(d). The $Ti_2N$ sub-layers preserve the zig-zag pattern separated by A atomic layers. The brightness of dots depends on the atomic mass (intensity~$Z^2$), which means that the heavier Cu atoms have replaced Al in the A layers (Figure 1(d)). The Cu layers are brighter than the Ti layers because of their difference in atomic mass. A lattice-resolved EDS mapping and line scan reveal the atomic positions in the crystal structure (Figure 1(e-f)). Cu is predominant in the final product and has the same atomic positions as Al (Figure 1(f)). The relative atomic ratio of (Al:Cu) at.% is about (1:9) at.% as identified by STEM-EDS, consistent with SEM-EDS results. Therefore, all above characterization results indicates that the Cu-incorporated MAX phase has a chemical formula of $Ti_2(Al_{0.1}Cu_{0.9})N$.

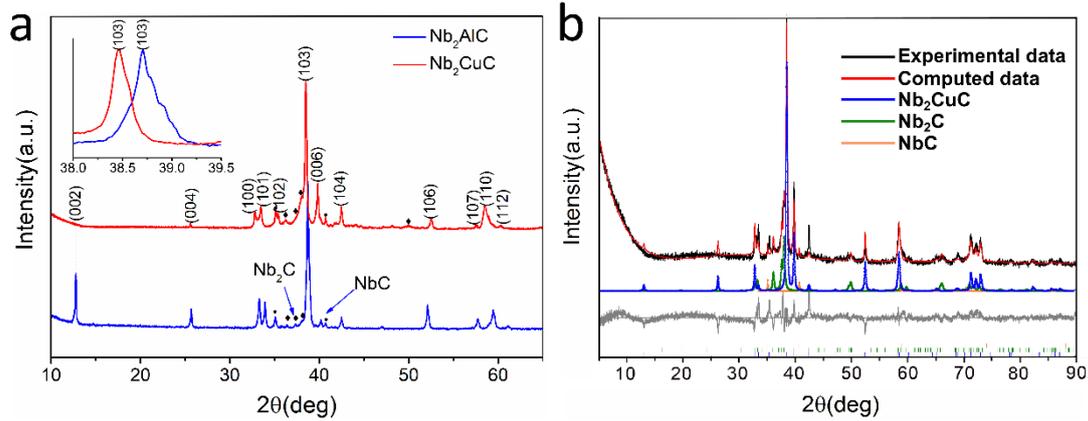

**Figure 2.** (a) XRD patterns of the Nb$_2$AlC and the Nb$_2$CuC obtained from the reaction of Nb$_2$AlC and CuI. (b) Rietveld refinement of XRD of the Nb$_2$CuC.

The low amount of Al in Ti$_2$(Al$_{0.1}$Cu$_{0.9}$)N is noteworthy. In our recent work on Ti$_3$AlC$_2$, only partial substitution of Cu in Ti$_3$(Al$_{1/3}$Cu$_{2/3}$)C$_2$ was achieved. Thus, the same reaction in other MAX phases should be investigated to understand the underlying mechanism. Figure 2(a) shows XRD patterns of the raw Nb$_2$AlC and the final product obtained through the same replacement methodology in CuI molten salt. Before treatment by ammonium persulfate solution, the characteristic peaks of Cu (2θ≈43°, 2θ≈51° and 2θ≈75°) are detected (Figure S(3)), which indicates the generation of Cu metal during the replacement reaction between Al (derived from Nb$_2$AlC) and CuI. In addition, it can be observed that main the diffraction patterns of the Nb$_2$AlC MAX phase and product after Cu replacement are similar, but the (002) diffraction peak of the Cu-incorporated MAX phase became significantly weaker, as in the case of Ti$_2$(Al$_{0.1}$Cu$_{0.9}$)N. On the contrary, the (006) peak became stronger, which means that the Cu substitution in between Nb$_2$C layers changes the stacking of atoms perpendicular to *c* plane [6,7,11]. To confirm the lattice parameters of the resultant product, Rietveld refinement of the XRD pattern was carried out assuming phase-pure Nb$_2$CuC, as shown in Figure 2(b). The simulated pattern, with a reliability factor $R_{wp}$ of 7.88%, is in good agreement with the experimental data. The previously reported lattice parameters of the $Nb_2AlC$ are a = 3.106 Å and c = 13.888 Å [22], whereas the calculated lattice parameters of $Nb_2CuC$ are a = 3.153 Å and c = 13.587 Å. The atomic positions of $Nb_2CuC$ determined from the Rietveld refinement are listed in Table 1.

**Table 1**. Atomic positions in $Nb_2CuC$ determined from the Rietveld refinement.

| Element | x | y | z | Symmetry | Wyckoff symbol |
|---------|-----|-----|---------|----------|----------------|
| Nb | 1/3 | 2/3 | 0.59272 | 3m | 4f |
| Cu | 1/3 | 2/3 | 0.25 | -6m2 | 2c |
| C | 0 | 0 | 0 | -3m | 4a |

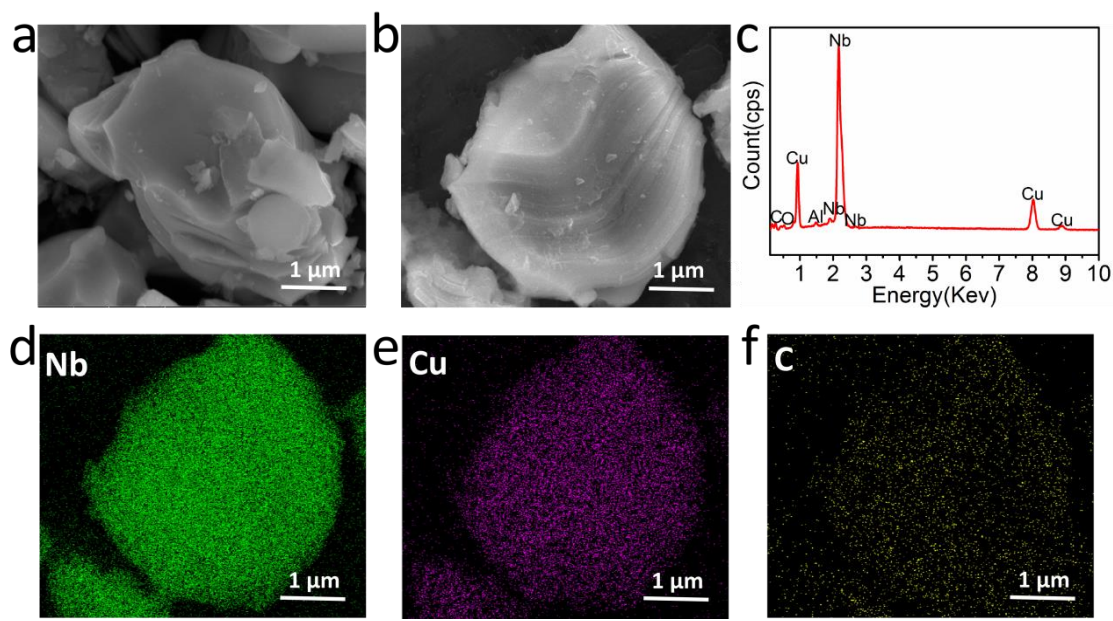

**Figure 3.** (a) SEM image of $Nb_2AlC$. (b) SEM image of the $Nb_2CuC$ obtained from the reaction between $Nb_2AlC$ and CuI. (c) EDS spectrum of (b). (d) - (f) EDS mapping of Nb-Lα, Cu-Kα, and C-Kα signals of (b).

SEM images of $Nb_2AlC$ and $Nb_2CuC$ particles are shown in Figure 3(a-b),

respectively. The Nb$_2$CuC retains the layered morphology like the raw Nb$_2$AlC. The EDS spectrum of Nb$_2$CuC is shown in Figure 3(c), and all Nb, Cu, and C elements were detected. Only a very small peak possibly originating from Al was seen in the EDS spectrum, and the Al content should thus be at a concentration near or below the detection limit. The relative atomic ratio of (Nb:Cu) at.% is about (2:1) at.%, which is equal to the stoichiometry of an M$_2$AX phase. Figure 3(d-f) illustrates the EDS element mapping results of Nb-Lα, Cu-Kα and C-Kα signals of the particle shown in Figure 2(b), respectively. The uniform distribution of Nb, Cu, and C in their respective atomic layers indicates that Cu has fully replaced Al in Nb$_2$AlC.

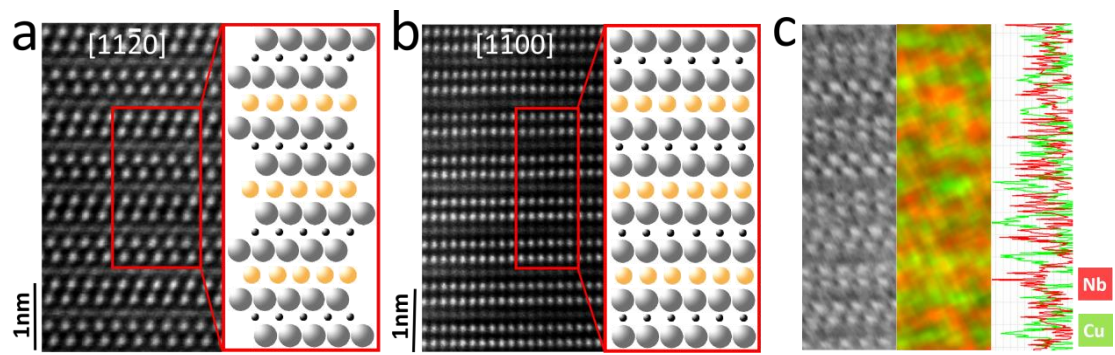

**Fig 4.** High-resolution (HR)-STEM images of Nb$_2$CuC showing atomic positions along [11$\bar{2}$0] (a) and [1$\bar{1}$00] (b) direction, respectively. (c) STEM-EDS mapping and line–scan of Nb-Lα (red) and Cu-Kα (green) signals, respectively.

In order to further determine the structure of Nb$_2$CuC, STEM was performed. Figure 4(a-b) show the atomic arrangements with the beam aligned along the [11$\bar{2}$0] and [1$\bar{1}$00] directions, respectively. As can be observed in both images, one darker layer (the A elements) interleaves two adjoining brighter layers of Nb with larger atomic mass. The presented images are similar to STEM images of other M$_2$AX phases having the characteristic zig-zag stacking of the 211 M$_{n+1}$X$_n$ layers [6,12]. In Figure 4(c), the atomic-resolved EDS element mapping and line-scan of Nb-Lα and Cu-Kα further identified the atom position of Cu, all at A sites, corroborating the synthesis of Nb$_2$CuC

MAX phase through a replacement reaction, to form a MAX phase with only Cu on the A site.

**Table 2** Lattice parameters (Å), elastic constants (GPa) and cleavage energy (J/m$^2$) of Ti/Nb (M) and Al/Cu (A) atomic layers.

| MAX phase | $a$ | $c$ | $C_{11}$ | $C_{33}$ | $C_{44}$ | B | G | cleavage energy (M-A) |
|---|---|---|---|---|---|---|---|---|
| Ti$_2$AlN | 2.996 | 13.627 | 296 | 273 | 125 | 153 | 112 | 2.147 |
| Ti$_2$CuN | 2.978 | 13.169 | 255 | 193 | 27.2 | 112 | 56.7 | 1.654 |
| Nb$_2$AlC | 3.127 | 13.896 | 338 | 293 | 139 | 175 | 124 | 2.287 |
| Nb$_2$CuC | 3.134 | 13.276 | 271 | 295 | 23.1 | 183 | 44.8 | 1.564 |

The structures of Ti$_2$AlN, Ti$_2$CuN, Nb$_2$AlC, and Nb$_2$CuC MAX phases were caculated by DFT calculations. The lattice parameters, elastic constants and cleavage energies of Ti/Nb and Al/Cu atomic layer in the MAX phases are listed in Table 2. The calculated lattice parameters of Ti$_2$CuN and Nb$_2$CuC show reduced $c$ values compared to Ti$_2$AlN and Nb$_2$AlC, respectively, consistent with our experimental results. However, the experimental $a$ value of Ti$_2$(Al$_{0.1}$Cu$_{0.9}$)N conflicts with the calculated $a$ value, which may be due to the fact that A layers are not completely replaced. The large decrease in elastic constant $C_{44}$ and shear modulus G indicate enhanced plasticity in the Cu-containing MAX phases. The phonon dispersion relations are plotted in Figure S4. The vibrational frequencies have no imaginary component, showing that all these MAX phases are dynamically stable. The variation of cleavage energy in the M-Al and M-Cu atomic layers suggests weaker bonding between M and Cu atoms, which implies that these Cu-containing MAX phases may be useful as precursors for two-dimensional MXene materials.

The incorporation of transition elements (Zn, Cu) into A site of MAX phase has been discussed in our previous reports where chloride salts were used [10,14]. Here, we used an alternative molten salt CuI, which has a melting point of 600°C. At 700°C, CuI is molten and ions of $Cu^+$ and $I^-$ can contact solid reactants [23]. As a strong electron acceptor or Lewis acid [24,25] in molten salt, $Cu^+$ oxidizes Al a drive it out from $Nb_2AlC$. The $Al^{3+}$ cation is then coordinated with $Cl^-$ to form $AlCl_3$ which evaporates (boiling point ~360°C). However, the occupancy of copper in the final products is not simply determined by the phase diagram. In the Au-Si and Al-Zn binary phase diagrams, two end members of metal components are separated below the solidus line, which provided a predictable guideline for synthesis of new MAX phases, such as $Ti_3AuC_2$ and $M_{n+1}ZnX_n$ (M=Ti or V; X=C or N, n=1 or 2) [7,10]. In contrast, based on the Cu-Al binary phase diagram, it does not appear possible to obtain a single-phase Cu end member from the Al-rich side, since equilibrium intermediate Al-Cu alloys should form, which is why $Ti_3(Al_{1/3}Cu_{2/3})C_2$ was formed in earlier work [14]. Previous reports also indicated this alloying behavior of Cu into A site (Si [26] or Al [27] elements). However, the present work indicates that this reasoning based on phase-diagram-guidelines is incomplete, since an end member with only Cu on A sites ($Nb_2CuC$) and one with very high Cu content ($Ti_2(Al_{0.1}Cu_{0.9})N$) were formed. The atomic arrangement in bulk materials, accurately predicted in phase diagrams, will change in a confined space due to the different crystal field strength exerted by other components. In the nanolaminated MAX-phase crystal structure, A atoms have relatively weak bonding with the nearest M atoms and negligible bonding with the next A layer. When an A′ atom (here Cu) occupies the original A-atom position, the crystal field exerted by M atoms modifies the arrangement of A′ in A atomic sites, which may explain why full replacement of A atoms can be achieved in the present work despite the thermodynamic tendency to form a Cu-Al mixture.

## Conclusion

In summary, the new MAX phases of $Ti_2(Al_{0.1}Cu_{0.9})N$ and $Nb_2CuC$ were synthesized by A-site replacement reaction in molten salt environment. Complete or partial occupancy of Cu at the original Al site in the final MAX phases was demonstrated through atomically-resolved scanning transmission electron microscopy. Density-functional theory calculations corroborated the structural stability of these new MAX phases and predicted their elastic properties and low cleavage energy of M-A bonding, suggesting these Cu-containing MAX phases as precursors for future two-dimensional MXene derivation.

**AUTHOR INFORMATION**
**Corresponding author**
*Email: huangqing@nimte.ac.cn

†These authors contributed equally to this work.

**Conflict of interest**: The authors declare no competing financial interests.

**Acknowledgements**

This study was supported financially by the National Natural Science Foundation of China (Grant No. 21671195 and 91426304), Chinese Academy of Sciences (grant No. 2019VEB0008 and 174433KYSB20190019), Swedish Government Strategic Research Area (No. 2009 00971), the Knut and Alice Wallenberg Foundation (Grant KAW 2015.0043), Swedish Foundation for Strategic Research (EM16 - 0004) and the Research Infrastructure Fellow (RIF 14-0074).